\documentclass{article}

\usepackage{microtype}
\usepackage{graphicx}
\usepackage{subfigure}
\usepackage{booktabs} 
\usepackage{todonotes} 

\usepackage{hyperref}

\usepackage[accepted]{mlsys2025}
\usepackage{amsmath}

\mlsystitlerunning{MonoMoE: An Efficient Fused Mega-kernel for Quantized MoE Decoding}

\begin{document}

\twocolumn[
\mlsystitle{MonoMoE: An Efficient Fused Mega-kernel for Quantized MoE Decoding}




\begin{mlsysauthorlist}
\mlsysauthor{Yu Gong}{amz}
\mlsysauthor{Kailash Budhathoki}{amz}
\mlsysauthor{Taeho Kim}{amz}
\mlsysauthor{Haipeng Li}{amz}
\mlsysauthor{Ashish Khetan}{amz}
\end{mlsysauthorlist}

\mlsysaffiliation{amz}{Amazon AGI, Santa Clara, USA}

\mlsyscorrespondingauthor{Yu Gong}{ygkyle@amazon.com}

\mlsyskeywords{Machine Learning, MLSys}

\vskip 0.3in

\begin{abstract}
Mixture-of-Experts (MoE) layers increase model capacity without proportionally increasing arithmetic, but their sparse expert computation is difficult to execute efficiently during autoregressive decode. Existing grouped and batched GEMMs are token-major: they construct expert-local token tiles and obtain parallelism from the token dimension. When few tokens reach each expert, this organization incurs tile padding and preprocessing, launches short-lived grids that underutilize memory bandwidth, and exposes quantization, activation, and reduction as separate stages. We present \textbf{MonoMoE}, a weight-major persistent megakernel for block-wise quantized MoE decode. MonoMoE places the complete decode-step token tile on the fine-grained tensor-core $N$ dimension and partitions CTAs over expert-weight tiles, eliminating expert-local token materialization and reducing padded arithmetic. A persistent grid fuses routing, top-$k$ selection, quantization, both expert projections, activation, and reduction in one launch; warp specialization and readiness flags overlap auxiliary work with the dominant expert-weight stream. MonoMoE is integrated with vLLM and supports multiple model shapes through generated kernel specializations and offline schedule tuning. On NVIDIA H200 GPUs, MonoMoE accelerates the complete routed-MoE operator by up to $\mathbf{1.54\times}$ over vLLM Triton Grouped GEMM, is $\mathbf{2.20}$--$\mathbf{3.84\times}$ faster than FlashMoE-FP8 adaptation across various models and batch sizes, and reduces end-to-end time per output token by up to $\mathbf{18.7\%}$ across the evaluated FP8 models, while preserving task accuracy. The MonoMoE implementation and supporting artifacts are open source and available in the \href{https://github.com/flashinfer-ai/flashinfer/tree/main/csrc/fused_moe/monomoe}{FlashInfer repository}.
\end{abstract}
]



\printAffiliationsAndNotice{}  

\section{Introduction}
\label{submission}

Mixture-of-Experts (MoE) models scale capacity by replacing each dense feed-forward network (FFN) with a set of sparsely activated experts, only a few of which process each token~\cite{shazeer2017moe,lepikhin2021gshard,fedus2022switch,du2022glam,jiang2024mixtral}. This design increases model quality and robustness without proportionally increasing per-token arithmetic. Sparse activation, however, does not make decode cheap: in our vLLM measurements~\cite{kwon2023vllm}, routed MoE layers account for up to 40\% of end-to-end decoding latency. Each decode step performs little computation per selected expert, yet must fetch the corresponding expert weights from HBM, making expert execution memory-bound despite its modest FLOP count. Quantized weights and activations reduce this traffic~\cite{micikevicius2022fp8,xiao2023smoothquant}, but routing, activation quantization, and scale handling remain exposed on the critical path.

Current GPU MoE kernels commonly express expert computation as grouped or batched GEMM~\cite{he2021fastmoe,hwang2023tutel,rajbhandari2022deepspeedmoe,gale2023megablocks}. These implementations are fundamentally \emph{token-major}: routed tokens form the GEMM $M$ dimension, and work is partitioned into expert-local token and output tiles. This organization is effective for training, prefill, and large-batch serving, where each expert receives enough tokens to fill its tiles and expose sufficient parallelism. Decode presents the opposite regime. A small number of tokens is distributed across many experts, leaving token tiles largely empty and requiring the runtime to construct expert-local work lists for short, low-concurrency grids. Moreover, the MoE forward pass is divided into separate routing, quantization, gate/up projection, activation, down-projection, and reduction kernels. Each kernel boundary interrupts the expert-weight stream and exposes auxiliary work that could otherwise be overlapped with weight movement.

Kernel fusion reduces launch overhead but does not essentially address the limitations of token-major scheduling. FlashMoE~\cite{aimuyo2025flashmoe}, for example, fuses dispatch, expert computation, and reduction into a single persistent kernel — but it still breaks activations into expert-local tiles and pulls its parallelism from the token dimension. When token counts are low, i.e., as during decoding, those tiles end up underfilled, and activations still have to be written out on the critical path. On top of that, FlashMoE's fixed SM scheduling budget flat-out refuses to run certain wide-model or higher-token configurations. So while FlashMoE does a better job overlapping stages and hiding communication costs, it still inherits the same token-major scheduling assumptions that make decode efficiency hard in the first place.

Our profiling puts numbers to this inefficiency. For a single Qwen3.5-35B-A3B-FP8 layer at batch size one on an H200, vLLM spends 23.0\,$\mu$s in the two expert GEMMs -- yet these kernels sustain only 17--22\% of peak DRAM bandwidth. Routing, quantization, activation, and reduction add a comparable amount of low-bandwidth work on the same critical path. Across decode token counts from one to eight, our entitlement analysis shows that closing kernel-boundary gaps and recovering memory throughput can cut inter-token latency by 32--36\%. The expert-weight traffic itself is unavoidable — but repeatedly interrupting that traffic and serializing auxiliary work around it does not have to be.

We build \textbf{MonoMoE} guided by a different principle. MonoMoE is \emph{weight-major}: it partitions expert weights into output-row stripes and places the complete decode-step token tile on the tensor core's fine-grained $N$ dimension. Parallelism now comes from the weight matrices rather than from expert-local token blocks, so the GPU has useful work even when an expert receives only one or two tokens. This mapping removes off-chip token sorting and activation dispatch, reduces padded tensor-core work, and keeps the original token tile and compact routing state on chip.

MonoMoE realizes this mapping as one persistent CUDA kernel. A fixed resident grid iterates across selected experts while fusing router-logit computation, top-$k$ selection, activation quantization, gate/up projection, nonlinear activation, down projection and reduction. Static warp specialization keeps tensor-core math separate from asynchronous data movement, and fine-grained readiness flags connect producers to consumers without launch boundaries or grid-wide phase barriers. The net effect is that expert weights stay in flight while input preparation and epilogues run in the overlap windows this creates. From there, generated kernel specializations and offline tuning adapt the same dataflow to different model shapes, tensor-parallel degrees, and routing policies.

We integrate MonoMoE into vLLM~\cite{kwon2023vllm} and evaluate four block-wise FP8 MoE configurations on H200 GPUs. Over the same routed-MoE operator, MonoMoE is up to $1.54\times$ faster than vLLM Triton Grouped GEMM. Compared with the FlashMoE-FP8 adaptation, it is $2.20$--$3.84\times$ faster across various models and batch sizes. Unlike FlashMoE, MonoMoE executes every evaluated serving configuration without a fixed expert-plus-dispatch SM-capacity limit. In end-to-end serving, MonoMoE reduces mean time per output token by 9.9--18.7\% on Qwen3.5-35B, 9.1--11.5\% on Qwen3.5-122B, and 2.9--16.7\% on GLM-5.2. DeepSeek-V3.1's eight-way tensor-parallel path dilutes the local-kernel gain: TPOT ranges from a 1.2\% regression to a 3.7\% reduction, while aggregate output throughput improves by up to 31.8\%. GSM8K and HumanEval scores remain within the reported uncertainty whenever MonoMoE's mean is below the baseline.

In summary, we make three contributions:
\begin{itemize}
    \item We characterize why token-major MoE execution is inefficient during low-token decode and quantify the latency available from removing kernel boundaries and recovering memory throughput.
    \item We introduce a weight-major MoE decomposition and a persistent kernel that fuses the complete quantized routed path while overlapping auxiliary work with expert-weight movement.
    \item We integrate generated shape specialization and offline schedule tuning into vLLM, and demonstrate kernel-level and end-to-end gains across multiple model shapes and tensor-parallel configurations without degrading task accuracy.
\end{itemize}

\section{Bottleneck Analysis \& Related Work}
\label{bottleneck}   

We begin with a conservative \emph{entitlement analysis} that quantifies the optimization opportunity in MoE decode. The analysis estimates \emph{how much} latency is recoverable from the multi-kernel baseline, distinguishes losses due to throughput gap from kernel boundary, and identifies the \emph{batch-size regime} in which the opportunity is largest. We then trace this opportunity to the structural limitations of token-major execution. Together, these findings motivate the weight-major persistent design of MonoMoE in Section~\ref{design}.

\begin{figure*}[ht]
    \centering
    \includegraphics[width=\linewidth]{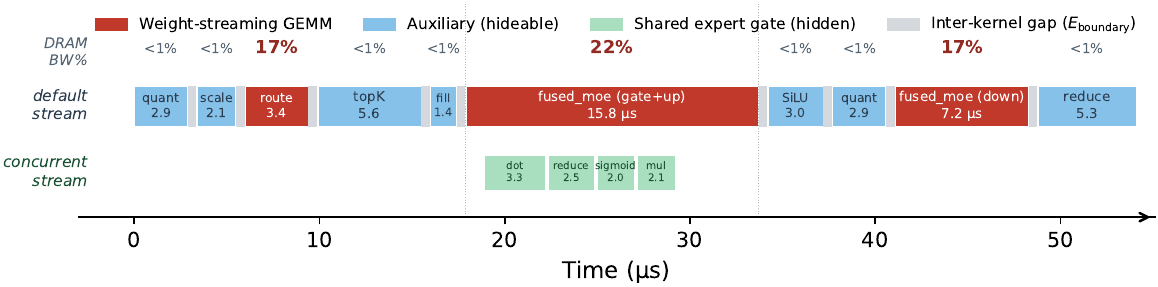}
    \vspace{-1.5em}
    \caption{Kernel timeline for one MoE layer during decode at BS=1 (Qwen3.5-35B-A3B-FP8, H200, vLLM with Triton \texttt{fused\_moe} under \texttt{torch.compile} and CUDA graphs). The two weight-streaming GEMMs dominate wall-clock time yet sustain only 17--22\% of peak DRAM bandwidth, while all auxiliary kernels operate below 1\% and are fully hideable under a continuous weight stream. The shared-expert gate executes on a concurrent stream, entirely overlapped with the gate+up projection.}
    \label{fig:moe-breakdown}
\end{figure*}

\subsection{Entitlement: Quantifying the MonoMoe Kernel Opportunity}
\label{bottleneck:entitlement}
Under vLLM with \texttt{torch.compile} and CUDA graphs, a single MoE layer's routed path executes as several independently launched kernels spanning input quantization, routing, expert projections, activation, and reduction. For this analysis, we define a \emph{monokernel} as an idealized single persistent kernel that covers the same routed-MoE operator boundary and performs exactly the same computation as this baseline. It differs only \emph{how} and \emph{when} that work is scheduled on the GPU, not \emph{what} is computed. We decompose the recoverable latency into two additive, non-overlapping terms:
\begin{equation}
\label{eq:entitlement}
 E_{\text{total}} = E_{\text{boundary}} + E_{\text{throughput}}.
\end{equation}

\paragraph{$E_{\text{boundary}}$: inter-kernel idle time.}
$E_{\text{boundary}}$ is the measured GPU-idle time between consecutive kernel launches within each layer's MoE block---time during which no SM
executes useful work:
\begin{equation}
\label{eq:eboundary}
 E_{\text{boundary}} = \sum_{l=1}^{L} \sum_{i \in \text{block}_l}
   \max\!\bigl(0,\; t_{\text{start}}^{(i+1)} - t_{\text{end}}^{(i)}\bigr),
\end{equation}
where $L$ is the number of decoder layers and $\text{block}_l$ is the contiguous span of kernels launched in layer~$l$'s MoE forward pass, identified via \texttt{nsys} profiling under CUDA-graph replay. Despite CUDA-graph capture eliminating host-side launch overhead, the GPU still observes measurable idle intervals at kernel boundaries: each kernel's final warps must drain before the next kernel's warps begin filling the SMs, creating brief periods of partial or zero SM occupancy that accumulate across the several launches per layer. Under CUDA graphs, these gaps are small relative to total kernel runtime. The dominant opportunity lies within the kernels themselves -- the bulk of recoverable latency comes from bandwidth underutilization, which is captured by our second term.


\paragraph{$E_{\text{throughput}}$: bandwidth recovery within kernels.}
$E_{\text{throughput}}$ is the latency recovered by raising sustained HBM bandwidth from the baseline's achieved level to the monokernel's steady-state rate.  A persistent kernel that never tears down its memory pipeline sustains higher bandwidth than a sequence of short-lived launches, each of which independently pays startup and drain transients.
Formally:
\begin{equation}
\label{eq:ethroughput}
 E_{\text{throughput}} = \max\!\bigl(0,\; T_{\text{baseline}} - T_{\text{mono}}\bigr),
\end{equation}
where $T_{\text{baseline}}$ is the sum of kernel-active durations within the MoE blocks (excluding inter-kernel gaps, which belong to
$E_{\text{boundary}}$). $T_{\text{mono}}$ is the monokernel's estimated execution time. Because decode is memory-bound, the critical path is routing followed by weight streaming --- remaining work (quantization, top-$k$ selection, activation, and reduction) largely overlaps with weight loading: 

%
\begin{equation}
\label{eq:tmono}
 T_{\text{mono}} = T_{\text{routing}} +
   \frac{W}{\textrm{BW}_{\text{mono}}},
\end{equation}
where $W$ is the total expert weight bytes streamed per decode step (determined by the number of unique experts among $E$ experts activated across $L$ layers): 
\begin{equation}
\label{eq:brouted}
W = L \times U(E, K, B) \times (H \times 2I + I \times H) \times b,
\end{equation}
with $U(E, K, B) = E \times (1 - (1 - K/E)^{B})$ the expected number of unique experts activated per layer under uniform routing of $B$ tokens ($B$ is also the batch size during decode), $K$ the top-K routing parameter, $H$ the hidden dimension, $I$ the per-expert intermediate size, and $b = 1$ byte (FP8).
We estimate the monokernel's sustained bandwidth during weight streaming conservatively:
\begin{equation}
\label{eq:bwmono}
 \text{BW}_{\text{mono}} = \min\!\bigl(\beta_{\text{target}} \times
   \text{BW}_{\text{peak}},\;
   \text{BW}_{\text{baseline}} \times \alpha_{\text{max}}\bigr),
\end{equation}
with $\beta_{\text{target}} = 0.8$ (the empirical ceiling for sustained streaming on Hopper, reflecting irreducible losses from DRAM refresh, controller overhead, and access-pattern inefficiencies), and $\alpha_{\text{max}} = 1.5$ (a conservative cap on improvement over baseline through persistent residency and continuous pipelining). At low batch size, $\alpha_{\text{max}}$ binds because the baseline operates far below the hardware ceiling; at high batch size, $\beta_{\text{target}}$ binds because the baseline already approaches saturation.

\begin{table}[t]
\centering
\caption{Monokernel entitlement for Qwen3.5-35B-A3B-FP8 ($E{=}256$,
top-$k{=}8$) on H200, serving with vLLM (1600 input, 600 output tokens). 
$\text{BW} (\%)$ is the achieved fraction of peak HBM bandwidth in the baseline \texttt{fused\_moe} triton kernel.}
\label{tab:entitlement}
\vspace{0.5em}
\small
\begin{tabular}{rrrrrr}
\toprule
& \multicolumn{2}{c}{Baseline} & \multicolumn{3}{c}{Entitlement} \\
\cmidrule(lr){2-3} \cmidrule(lr){4-6}
BS & ITL (ms) & $\text{BW}$ (\%) & $E_{\text{bnd}}$ (ms) & $E_{\text{thr}}$ (ms) & \%ITL \\
\midrule
1  & 4.32 & 20.0 & 0.24 & 1.14 & 31.9 \\
4  & 5.07 & 46.1 & 0.39 & 1.41 & 35.5 \\
8  & 6.45 & 62.8 & 0.20 & 1.90 & 32.7 \\
16 & 8.55 & 60.1 & 0.34 & 1.10 & 16.8 \\
32 & 10.85 & 66.9 & 0.26 & 0.62 & 8.0 \\
\bottomrule
\end{tabular}
\end{table}
\paragraph{Measured entitlement.} Table~\ref{tab:entitlement} reports the entitlement for Qwen3.5-35B-A3B-FP8 (256 experts, top-8) on a single H200 (4.8\,TB/s peak HBM3e) at 1600 input and 600 output. At B$=$1--8, the monokernel can recover 32--36\% of inter-token latency. At B$=$32 only 8\% remains---the baseline GEMMs already sustain $\sim$67\% of peak. These results identify \textbf{small-batch decode} as the regime with the greatest remaining optimization headroom and thus the primary target for improved kernel execution.

Figure~\ref{fig:moe-breakdown} decomposes one MoE layer execution at B$=$1. The two \texttt{fused\_moe} GEMMs (gate+up: 15.8$\mu$s; down: 7.2$\mu$s) sustain only $\leq$22\% of peak DRAM bandwidth despite dominating layer time. The remaining $\sim$28\,$\mu$s of auxiliary kernels consume near-zero bandwidth and are fully hideable under a continuous weight stream. The opportunity is twofold: inter-kernel gaps contribute $E_{\text{boundary}}$, while the low sustained bandwidth of the short-lived GEMMs contributes $E_{\text{throughput}}$.

\begin{figure*}
    \centering
    \includegraphics[width=\linewidth]{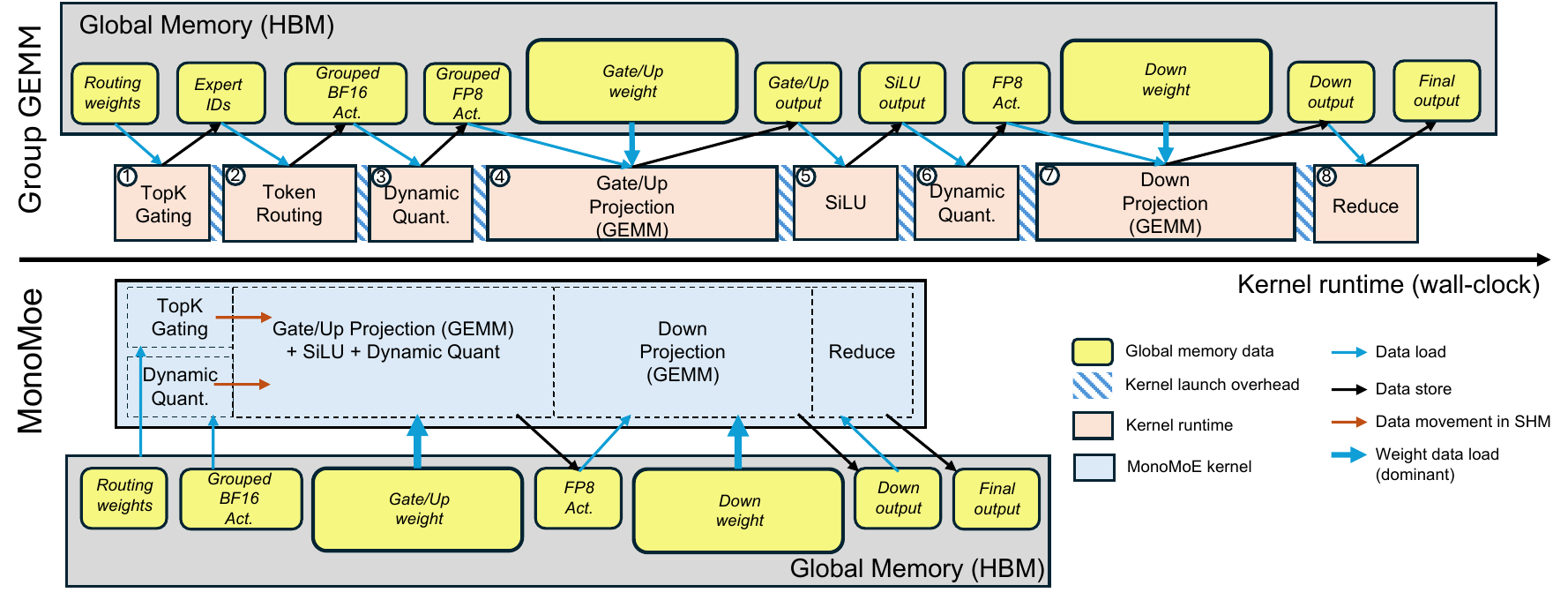}
    \vspace{-1.5em}
    \caption{Token-major grouped GEMM carves work from padded token and output tiles, then executes the two expert projections as separate stages. MonoMoE instead places tokens on the MMA's small axis, carves its persistent grid along expert-weight tiles, and keeps routing state on chip across the fused projections.}
    \label{fig:monoMoe_vs_grouped_gemm}
\end{figure*}

\subsection{Sources of the Deficit}
\label{bottleneck:sources}
The entitlement gap above originates from a common design choice in standard MoE kernels: they are \emph{token-major}~\cite{he2021fastmoe,hwang2023tutel,gale2023megablocks}. For each expert, routed tokens form the GEMM $M$ dimension, which is partitioned into \texttt{BLOCK\_M}-row tiles; the output dimension is partitioned independently into \texttt{BLOCK\_N}-column tiles. Grouped GEMM flattens the expert-local $M$ tiles into one work list, whereas batched (3D) GEMM retains an explicit expert dimension and masks each expert by its token count. Both designs derive a major source of parallelism from the token axis: with many tokens, they expose enough $M$ tiles to occupy the GPU, and the flattened grouped schedule additionally smooths imbalance across experts. Decode provides the opposite regime, exposing three costs that this decomposition poorly amortizes.

\subsubsection{Token Preprocessing Overhead}
\label{bottleneck:token}

\begin{figure}[t]
  \centering
  \includegraphics[width=\linewidth]{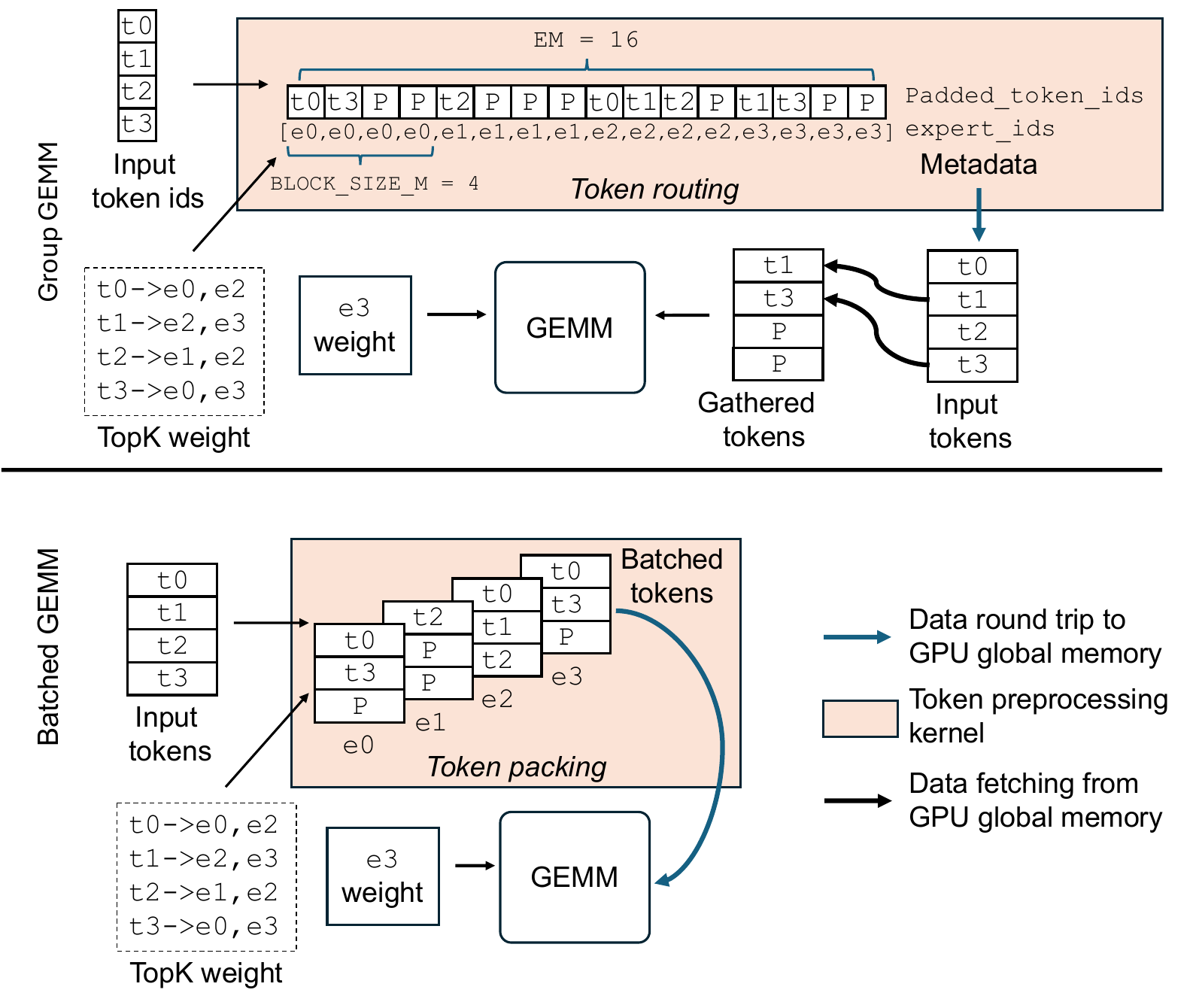}
  \vspace{-1.5em}
  \caption{Token-major preprocessing for grouped and batched GEMMs. Metadata-based grouped GEMM pads expert-local token-index lists, whereas contiguous grouped and batched layouts materialize expert-ordered token features. In all cases, the expert-local token dimension is rounded to the GEMM $M$-tile granularity.}
  \label{fig:token_preprocessing}
\end{figure}

Constructing a token-major GEMM requires converting the token -- expert assignments into expert-local rows. If expert $e$ receives $B_e$ tokens, its scheduled row count is rounded to the $M$-tile granularity, giving
\begin{equation}
\label{eq:token_major_padding}
M_{\mathrm{sched}}
=
\sum_{e\in\mathcal{A}}
\left\lceil\frac{B_e}{\texttt{BLOCK\_M}}\right\rceil
\times \texttt{BLOCK\_M},
\end{equation}
where $\mathcal{A}$ is the set of active experts. In decode, $B_e$ is commonly much smaller than \texttt{BLOCK\_M}, so most scheduled row positions are padding. This padding is universal to the token-major decomposition, but its materialization depends on the implementation.

The routing distribution affects the two layouts differently. Let $R=Bk$ denote the number of routed token -- expert pairs for routing fanout $k$. For grouped GEMM, flattening mitigates load imbalance across CTAs, but padding is maximized when the pairs are dispersed across as many experts as possible. In the decode regime where $R\leq E$, the worst case assigns one pair to each of $R$ experts, yielding $M_{\mathrm{sched}}=R\cdot\texttt{BLOCK\_M}$ and a useful-row fraction of only $1/\texttt{BLOCK\_M}$. For example, $B=8$, $k=8$, and \texttt{BLOCK\_M}$=16$ produce 64 useful rows among 1,024 scheduled rows, or 93.75\% padding. Batched GEMM suffers the opposite extreme: if all tokens select the same $k$ experts, then $M_{\mathrm{cap}}=B$ while only $k$ of the $E$ expert slices contain data. Its useful-row fraction is therefore $k/E$---3.125\% for $E=256$ and $k=8$---even before any additional tile-alignment padding.

As illustrated in Figure~\ref{fig:token_preprocessing}, Grouped GEMM materializes \texttt{sorted\_token\_ids} and one expert id per $M$ block. Padded entries are invalid indices, so the GEMM gathers valid rows directly from the original activation tensor and masks the unused lanes; it does \emph{not} copy a dense padded activation matrix. DeepGEMM's contiguous grouped path instead permutes the quantized token features into an expert-contiguous $[M_{\mathrm{sched}},H]$ buffer and records the expert assignment through \texttt{m\_indices}. Batched (3D) GEMM similarly dispatches real token features into an expert-major tensor $[E,M_{\mathrm{cap}},H]$, where $M_{\mathrm{cap}}$ is the per-expert capacity, and masks rows beyond each expert's token count. The latter two paths therefore incur a feature-sized HBM round trip, while the Triton path incurs a smaller metadata round trip and irregular activation gathers. All three expose token preprocessing before the expert-weight stream and waste tensor-core lanes on padded $M$ rows.

\subsubsection{Low Effective Memory Throughput}
\label{bottleneck:memory}

\begin{figure}[t]
    \centering
    \subfigure[Achieved DRAM throughput and launched CTA count for grouped GEMM as batch size increases.]{
        \includegraphics[width=0.86\linewidth]{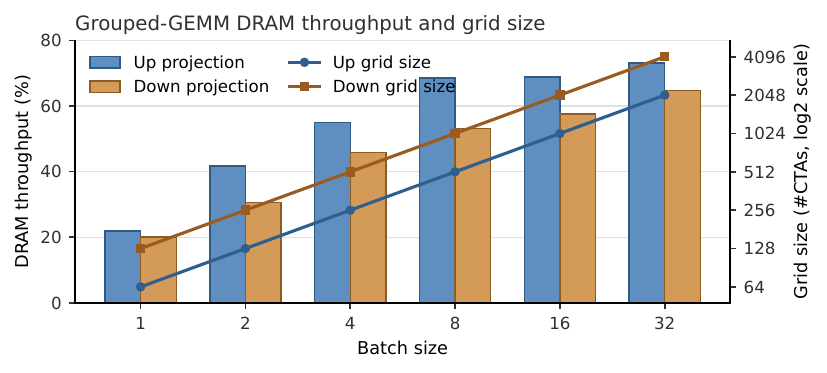}
        \label{fig:bw_histogram_vs_grid_size}
    }
    \vspace{-0.5em}
    \subfigure[Measured up-projection bandwidth trace of BS=1 up-projection kernel in Figure~\ref{fig:moe-breakdown}, showing that a short grouped-GEMM kernel spends most of its lifetime in fill and drain phases.]{
        \includegraphics[width=0.86\linewidth]{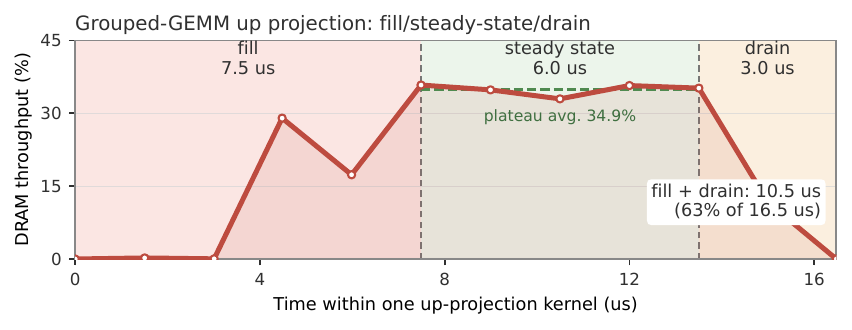}
        \label{fig:grouped_up_fill_drain}
    }
    \vspace{-0.5em}
    \caption{Grouped-GEMM DRAM throughput is limited by insufficient grid-level concurrency and memory-pipeline fill/drain transients.}
    \label{fig:memory_throughput}
\end{figure}

Although decode is dominated by weight movement, token-major GEMMs do not sustain peak HBM bandwidth in this regime. For an expert projection with output width $O$, the number of output tiles is
\begin{equation}
\label{eq:token_major_tiles}
T =
\sum_{e\in\mathcal{A}}
\left\lceil\frac{B_e}{\texttt{BLOCK\_M}}\right\rceil
\times
\left\lceil\frac{O}{\texttt{BLOCK\_N}}\right\rceil .
\end{equation}
Each CTA computes one $(M\text{-tile},N\text{-tile})$ output tile and streams the corresponding expert-weight panel. A flattened or persistent tile scheduler can balance these CTA tasks across SMs, but the token-major tiling fixes how many such tasks are available. When each active expert contributes only one mostly empty $M$ tile, the grid is determined primarily by the number of active experts and output-column tiles. The grid size can be too small or too short-lived to sustain enough outstanding HBM requests.

This effect appears as a discontinuous weight-load stream. The comes from the repeated startup, expert lookup or descriptor selection, and teardown around short tile sequences. The batched GEMM approach pipelines these loads with TMA, but a short schedule still provides little steady-state interval between pipeline fill and drain. The grouped GEMM exhibits the analogous behavior through its tiled global-memory load pipeline. As batch size grows, more experts become active or experts acquire additional $M$ tiles, increasing both grid-level concurrency and the duration over which weight requests remain in flight. This explains the throughput trend in Figure~\ref{fig:memory_throughput}.

Staging further degrades the memory throughput. Quantization, token routing, expert projections, SiLU activation, and reduction all fire as separate kernels, with each kernel's output written back to off-chip memory before the next one can pick it up. On top of that, each stage incurs a launch and a new fill-and-drain transient as blocks are scheduled for new kernels, issue their first data loads, and build enough in-flight memory requests to approach steady state. Conventional execution therefore repeatedly stops and restarts expert-weight streaming rather than maintaining it across the MoE layer.

\subsubsection{Untapped Latency-Hiding Opportunity}
\label{bottleneck:optimization}
The ideal lower bound for an MoE layer in the decode regime is the time to determine the selected experts and fetch their weights; the remaining work should be overlapped with that dominant stream whenever dependencies permit. Figure~\ref{fig:moe-breakdown} illustrates the gap in the baseline: at B$=$1, the two expert GEMMs spend 23.0\,$\mu$s streaming weights, while the default stream also exposes roughly 23\,$\mu$s of low-bandwidth auxiliary work, including quantization, top-$k$ selection, activation, and reduction. The 3.4\,$\mu$s routing kernel contributes to the routing lower bound, and the shared-expert gate is already hidden on a concurrent stream. The remaining auxiliary kernels consume little DRAM bandwidth, so they represent scheduling slack rather than unavoidable weight traffic.

Grouped and batched formulations leave this overlap largely unused because of the sequential kernel execution. This serialization is especially costly for block-wise quantized models, where activation quantization is small relative to the GEMM but visible when placed on the critical path. A fused design should instead keep the expert-weight stream live while issuing routing, quantization, activation, and epilogue work asynchronously around it. The goal is not to reduce unavoidable weight traffic, but to make observed layer latency approach the weight-fetch lower bound.

\subsection{Existing Works}
\label{bottleneck:existing}
Most prior MoE systems and kernels are designed primarily for training, prefill, or throughput-oriented serving at large effective batch sizes~\cite{he2021fastmoe,hwang2023tutel,rajbhandari2022deepspeedmoe, gale2023megablocks}. In these regimes, many routed tokens reach each expert, so token-major tiles are well populated and the costs of dispatch, padding,
and kernel boundaries are amortized. Consequently, the latency-sensitive autoregressive decode regime, in which each expert receives only a few tokens, remains comparatively underexplored.

Kernel fusion and IO-aware operator design have emerged as broader directions for GPU inference~\cite{dao2022flashattention}. ThunderKittens~\cite{spector2024thunderkittens} provides tile-level abstractions for constructing asynchronous, persistent GPU pipelines, while KOG's MI300X inference engine~\cite{kog2026singlekernel} executes an LLM through a single long-lived kernel. SonicMoE~\cite{guo2025sonicmoe} applies related techniques to grouped GEMM through TMA-based movement, IO--compute overlap, and tile-aware token rounding. These techniques reduce padding and improve utilization when experts receive sufficiently many tokens, but they do not resolve how sparse MoE work should be decomposed when only a few tokens reach each expert. SonicMoE retains token-major expert tiles and executes the forward projections and aggregation as separate stages; when per-expert token counts fall below a tile, padding, short-lived grids, and off-chip intermediates remain.

FlashMoE~\cite{aimuyo2025flashmoe} and DeepGEMM's Mega MoE~\cite{deepgemm2026megamoe} extend this trend to distributed MoE by consolidating dispatch, expert computation, and combine into a persistent kernel and overlapping communication with tensor-core execution. Nevertheless, their storage and scheduling remain token-major. FlashMoE packetizes selected token activations into expert-local communication buffers, while Mega MoE packs dispatched activations into expert-contiguous, \texttt{BLOCK\_M}-aligned pools and schedules $(m\text{-block},n\text{-block})$ tasks. This organization is effective when many routed tokens provide enough blocks to fill the machine and amortize dispatch; in low-token decode, it retains activation materialization and exposes underfilled token tiles. MonoMoE targets this complementary regime with a weight-major persistent kernel: it consumes the original token tile, performs top-$k$ selection on chip, places the complete token dimension on the MMA's small axis, and partitions work over expert-weight tiles.

\section{MonoMoE Design}
\label{design}

MonoMoE is designed around three bottlenecks identified in Section~\ref{bottleneck}: token-axis preprocessing, limited bandwidth utilization from short-lived GEMM grids, and unexplored overlap between auxiliary work and the expert-weight stream. Therefore, we propose a \emph{weight-major} decomposition: tokens occupy the small axis of each tensor-core operation, while the persistent grid is partitioned over expert-weight tiles. Rather than optimizing the conventional stages independently, MonoMoE executes the routed MoE layer as a single persistent megakernel for top-$k$ inference on modern tensor-core GPUs. The kernel is parameterized by the per-step token count $B$, number of experts $E$, routing fanout $\text{top-}k$, hidden dimension $\textsc{H}$, and the gate/up/down projection shapes.

As shown in Figure~\ref{fig:monoMoe_vs_grouped_gemm}, the layer is implemented as one \texttt{\_\_global\_\_} kernel with four logical stages: top-$k$/quantize, up-projection, down-projection, and reduce. Dependencies between stages are communicated through readiness flags: a consumer proceeds once its required producer output is available, without a separate kernel launch or a grid-wide phase barrier. This fine-grained handoff keeps the persistent grid active and allows independent work to advance as soon as its inputs are ready. The rest of this section describes how the kernel organizes tokens, persistent execution, and asynchronous overlap.

\subsection{Weight-Major, Padding-Free Token Handling}
\label{design:token}
As quantified in Section~\ref{bottleneck:token}, token-major kernels round each active expert's token rows to the $M$-tile granularity, underutilizing tensor-core lanes and introducing padded arithmetic. Modern GPU architectures, including Hopper and Blackwell, provide fine-grained tensor-core $N$ shapes in multiples of eight~\cite{nvidia2022hopper,nvidia2025ptx}. This asymmetry motivates MonoMoE's \emph{weight-major} decomposition.

Rather than partitioning the grid over expert-local token blocks and output-column blocks, MonoMoE partitions it over expert-weight tiles: each CTA owns a stripe of output rows, streams the corresponding weight panels, and applies them to the complete generation-step token tile. The token dimension is therefore an operand dimension within each CTA rather than a source of CTA-level work. As illustrated in Figure~\ref{fig:monoMoe_vs_grouped_gemm}, this mapping exposes parallelism from the large weight matrix even when an expert has too few tokens to form multiple $M$ tiles.

To reduce this underutilization, MonoMoE adopts SwapAB by placing tokens on the small-$N$ side of the MMA. Let $X_e$ of shape $B_e\times D$ denote the compact activations selected for expert $e$, and let $\widetilde{X}_e$ of shape $B\times D$ denote an expert-masked view of the original token tile. For expert weight $W_e$ of shape $D\times O$, the conventional and weight-major formulations are
\begin{equation}
\label{eq:swapab}
\begin{aligned}
Y_e = X_e W_e
&: (M,N,K)_{\mathrm{std}} = (B_e, O, D), \\
\widetilde{Y}_e^\top = W_e^\top \widetilde{X}_e^\top
&: (M,N,K)_{\mathrm{SwapAB}} = (O, B, D),
\end{aligned}
\end{equation}
where $D$ is the reduction dimension and $O$ is the expert output width. SwapAB exchanges the logical operand roles and produces a transposed output layout; it does not launch a transpose kernel or move an additional tensor through global memory. For FP8, Hopper WGMMA supports $N=8,\ldots,256$ in multiples of eight, and Blackwell's UMMA/\texttt{tcgen05} path provides analogous small-$N$ shapes. MonoMoE selects the smallest supported $N$ that covers the step's token count.

The finer $N$ granularity also reduces padded arithmetic. Let $N_{\mathrm{hw}}(B)$ be the selected hardware $N$ shape and let $\mathcal{A}$ be the active experts. Ignoring constant factors, the scheduled work for a projection with reduction width $D$ and output width $O$ is
\begin{equation}
\label{eq:weight_major_compute}
\begin{aligned}
C_{\mathrm{token}}
&\propto
\sum_{e\in\mathcal{A}}
\left\lceil \frac{B_e}{\texttt{BLOCK\_M}} \right\rceil
\texttt{BLOCK\_M}\,D O, \\
C_{\mathrm{weight}}
&\propto
|\mathcal{A}|\,N_{\mathrm{hw}}(B)\,D O.
\end{aligned}
\end{equation}
At the smallest supported shape, $N_{\mathrm{hw}}=8$. Relative to the token-major $\texttt{BLOCK\_M}=16$ baseline analyzed in Section~\ref{bottleneck:token}, this halves the scheduled token-side tensor-core work for an active expert. More generally, MonoMoE selects among wider $N$ shapes according to the decode-step token count. The computation savings depend on the routing distribution and are greatest when the selected $N$ shape is smaller than the aggregate token-major $M$-tile footprint.

Finally, the original input remains a dense $B\times H$ tile. Tokens are neither sorted by expert nor dispatched into a capacity-padded $[E,M,H]$ tensor. Each resident CTA loads the input tile on chip and uses compact routing metadata to select the token lanes relevant to its current expert. Unused MMA lanes are masked logically rather than materialized as padded token indices or activation rows, and experts with no assigned tokens are skipped entirely. Thus, \emph{padding-free} refers to routed activation storage: MonoMoE eliminates the off-chip permutation or dispatch round trip while also reducing, though not always eliminating, unused tensor-core lanes.

\subsection{Persistent Single-Launch Execution}
\label{design:persistent}

\begin{figure}[t]
  \centering
  \includegraphics[width=\linewidth]{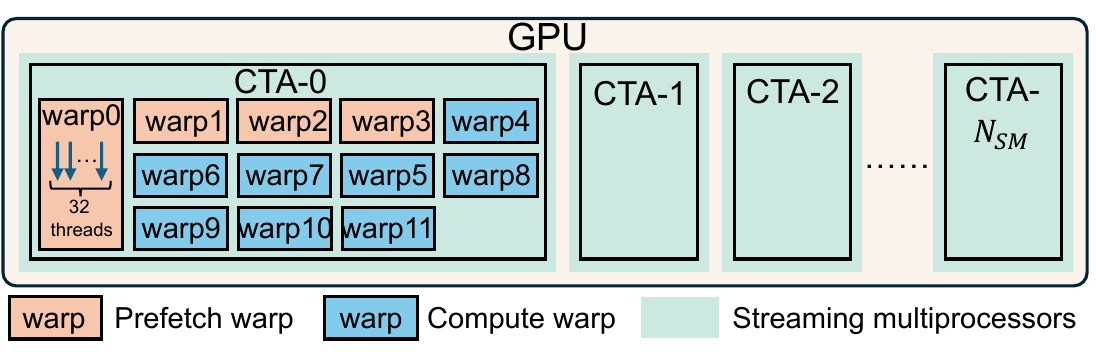}
  \vspace{-1.5em}
  \caption{MonoMoE persistent-grid mapping. The kernel launches one CTA per SM, so the grid contains $N_{\mathrm{SM}}$ CTAs on a GPU with $N_{\mathrm{SM}}$ streaming multiprocessors. Each CTA contains 12 warps: eight compute warps that issue tensor-core work and four prefetch warps that drive asynchronous data movement. With 32 threads per warp, each CTA has 384 threads.}
  \label{fig:gpu}
\end{figure}

\begin{figure}[t]
  \centering
  \includegraphics[width=0.8\linewidth]{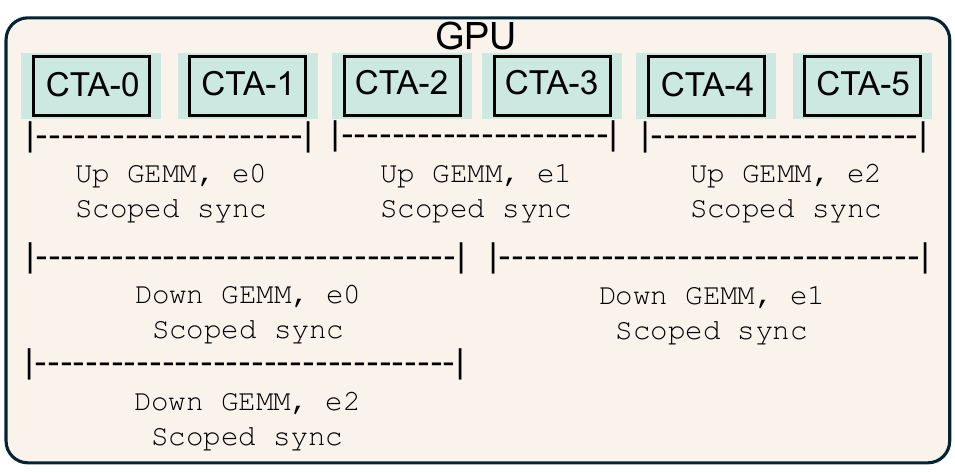}
  \vspace{-1em}
  \caption{Example MonoMoE persistent schedule with three experts and six CTAs. During the up-projection, CTAs form three two-CTA expert groups: CTAs in the same group stream the same expert's weights while computing disjoint output stripes. The groups then advance through experts in round-robin order. The down-projection uses a different grouping, with three CTAs per active expert, but follows the same pattern of expert-local weight streaming and stripe-wise output partitioning.}
  \label{fig:expert_parallel}
\end{figure}

Section~\ref{bottleneck:memory} shows that short grouped-GEMM grids repeatedly pay memory-pipeline fill and drain costs. MonoMoE changes the execution granularity: a fixed set of resident CTAs stays active for the full MoE layer, and each CTA streams work from multiple experts across both projections. This turns many short weight streams into a smaller number of long-lived streams, giving the memory system enough time and outstanding requests to approach steady state.

\paragraph{Persistent residency.}
MonoMoE launches at most one CTA per SM, as illustrated in Figure~\ref{fig:gpu}. The one-CTA residency is enforced by a launch bound of one block together with an opt-in dynamic shared-memory allocation. The host launcher caps the grid size by the SM count and verifies the realized occupancy at runtime. This residency guarantee is important for in-kernel synchronization. Every CTA is scheduled concurrently, so a consumer can safely wait on a readiness flag while its producer continues to run. The kernel remains a standard CUDA launch rather than a cooperative launch, so it can still be captured by CUDA Graphs.

\paragraph{Persistent expert streaming.}
MonoMoE groups CTAs only as an intra-kernel scheduling mechanism, not as a 3D-GEMM-style expert-major activation layout or a distributed expert-parallel partition. The CTAs in one group jointly cover the output range of one expert, then advance through the expert list with a fixed stride equal to the number of groups. A CTA therefore processes a sequence of experts over its lifetime, instead of retiring after one expert tile. MonoMoE chooses the CTA group size independently for the up and down projections, allowing each GEMM to use a partition suited to its output shape. Readiness flags connect these schedules: a down-projection group proceeds once its required up-projection data has been published, without requiring both projections to share the same CTA grouping. Within each projection, the same CTA owns the same output stripe for every expert it visits, so the scale layout, accumulator layout, and shared-memory tiling remain stable across the loop. Section~\ref{design:overlap} describes how prefetching carries the weight pipeline across expert boundaries, so the next expert begins with warm buffers rather than a cold memory pipeline.

\paragraph{Readiness-based handoff.}
Within the persistent launch, producer-consumer ordering is enforced through readiness flags rather than phase-wide barriers. A producer publishes a flag after completing the data required by the next stage, and a consumer waits only for the flag associated with its input. Independent expert groups can therefore progress at different rates instead of synchronizing at a common boundary. After the down-projection, partial sums are accumulated with \texttt{atomicAdd} into a single output buffer, and the corresponding readiness state determines when the final reduction can consume the completed output. This mechanism preserves data dependencies without introducing grid-wide synchronization.

\paragraph{Shared-memory reuse.}
Persistence is only practical if the live state of the fused layer fits within the per-SM shared-memory budget. MonoMoE uses a lifetime-aware shared-memory layout in which buffers with disjoint lifetimes alias the same storage. The routing-window input tile, the up-projection weight tile, and the down-projection weight tile share one region because their lifetimes do not overlap; similarly, epilogue scratch buffers are collapsed into a single reusable region. This reuse keeps the persistent footprint small enough to maintain one CTA per SM at the target tile sizes.

\subsection{Latency Hiding via Asynchronous Overlap}
\label{design:overlap}

\begin{figure*}
    \centering
    \includegraphics[width=\linewidth]{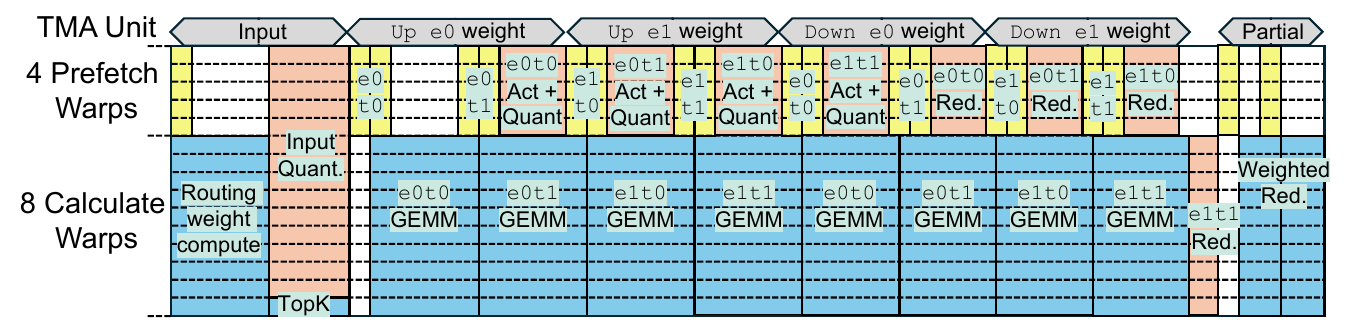}
    \vspace{-1.5em}
    \caption{MonoMoE schedule with dynamic warp specialization for a CTA traversing two experts. Labels use zero-based expert and tile indices; for example, \texttt{e0t0} denotes the first tile of expert~0. The four prefetch warps issue TMA transfers (narrow yellow
    markers) for future input, weight, and partial-output tiles while the eight calculation warps execute routing, top-$k$, and the current GEMM (blue). Activation and requantization after the up-projection, and partial accumulation after the down-projection, are deferred to the prefetch warps and overlapped with the next expert's GEMM. Only the final weighted reduction remains in the pipeline drain.}
    \label{fig:scheduling}
\end{figure*}

Asynchronous overlap turns the latency-hiding opportunity of Section~\ref{bottleneck:optimization} into a concrete schedule. Figure~\ref{fig:monoMoe_vs_grouped_gemm} provides the dataflow-level view: instead of separating token preparation and the two expert projections into independently launched stages, MonoMoE executes them within one persistent kernel. Since decode latency is lower-bounded by routing and expert-weight fetch, MonoMoE places the remaining work underneath the weight stream whenever dependencies allow. Accordingly, the persistent schedule overlaps routing, input quantization, matmul epilogues, and activation quantization for the next projection with data movement whenever their dependencies permit. This schedule uses dynamic \emph{warp specialization}, as illustrated in Figure~\ref{fig:scheduling}. Each CTA contains fixed groups of eight calculation warps and four prefetch/epilogue warps, but their work changes across pipeline phases: the calculation warps switch from routing and top-$k$ selection to tensor-core GEMMs, while the prefetch warps alternate among asynchronous data movement, scale loading, and deferred epilogues. Thread indices define the warp groups, while the pipeline state determines their current roles.

Figure~\ref{fig:scheduling} shows this division over two experts. Input
movement first overlaps routing and top-$k$ selection. During each projection, the prefetch warps issue TMA requests several iterations ahead while the calc warps consume the current weight tile. After expert~0 finishes itsup-projection, its activation and requantization execute concurrently with expert~1's up-projection; the down-projection uses the same handoff for partial accumulation. The narrow yellow markers denote the issue of an asynchronous transfer rather than its full duration: the transfer remains in flight across the following compute interval. This organization leaves only work without a successor, such as the final weighted reduction, exposed at pipeline drain.

\paragraph{Overlapping input fetch with router-logits computation.}
The first overlap window appears before expert weights are selected. Router-logit computation requires only the small router weights, while the full input tile is consumed later by the expert projections. MonoMoE therefore streams the input tile into shared memory asynchronously while the calc warps compute router logits. The input HBM read, which would otherwise be a serial prologue, is hidden behind routing and the tile is resident by the time quantization begins.

\paragraph{Overlapping metadata preparation with input quantization.}
After routing, one warp constructs the compact routing metadata while the remaining warps quantize the resident input tile in a strided partition. The quantized activation tile is single-buffered over the full reduction dimension, so the up-projection consumes it directly without alternating between quantization and compute. Together with the previous overlap, input loading, routing, top-$k$ selection, metadata construction, and input quantization are removed from the serial critical path: input loading hides behind routing, and metadata construction hides behind quantization.

\paragraph{Overlapping weight fetch with compute.}
Each projection uses a pipelined producer/consumer loop. Prefetch warps fetch future weight tiles while calc warps consume the current tile. MonoMoE uses multi-stage lookahead: buffers are armed several iterations before use, keeping multiple HBM transfers in flight and giving each transfer multiple compute windows in which to complete. This depth keeps the weight-load pipeline populated across iterations, producing a continuous weight stream and allowing the kernel to sustain higher DRAM bandwidth. It is especially important in the decode regime, where each tensor-core instruction consumes many more operand bytes than it produces output elements, making iteration time sensitive to transfer latency and L2 contention. The same lookahead extends across expert boundaries. During the final iterations of one expert, the prefetch warps load the first tiles of the next expert, so the next expert begins with warm buffers rather than a cold memory pipeline. The down-projection uses the same schedule, advancing weight and activation prefetches in lockstep.

\paragraph{Overlapping epilogues with the next expert.}
Once the prefetch warps issue TMA requests, the Tensor Memory Accelerator carries out the transfers asynchronously, leaving these warps idle until the next prefetch point. MonoMoE uses this available interval to execute fused per-expert epilogues, moving them off the calc-warp critical path. For the up-projection, the gate and up outputs, activation nonlinearity, output-scale reduction, and cast to the down-projection input are performed in registers and shared memory with a single global store, replacing the store/reload/requantize sequence of a staged implementation. MonoMoE executes this epilogue on the prefetch warps during the first iterations of the next expert's reduction loop. This hides the nonlinearity, special-function operations, reduction, and store behind the next expert's matmuls. A small routing-weight cache, populated at the barrier that publishes the matmul output, avoids a per-token scan that would otherwise compete with calc-warp shared-memory accesses. The down-projection follows the same pattern: accumulation into the output buffer is deferred to the prefetch warps during the next expert's first iteration. Only the final expert in a block's sequence, which has no successor to hide behind, runs its epilogue inline.

\subsection{Generic Kernel Interface and Shape-Specific Tuning}
\label{design:tuning}

MonoMoE separates algorithmic generality from hardware-aware specialization. All generated kernels implement the same conceptual operator interface:
\[
\begin{aligned}
\operatorname{MonoMoE}_{E,H,N,B_{\max},s}
(&X,R,W_{\mathrm{gu}},W_{\mathrm{down}},S;\\
 &B,k,p),
\end{aligned}
\]
where $X$ is the input token tile, $R$ denotes the routing inputs, $W_{\mathrm{gu}}$ and $W_{\mathrm{down}}$ are the expert weights, $S$ contains their block scales, $p$ is the routing policy, and $s$ is the selected schedule. A common kernel template implements routing, activation quantization, both expert projections, nonlinear activation, and weighted reduction. The token count $B$, routing fanout $k$, and routing policy $p$ remain runtime parameters, whereas the per-GPU model shape $(E,H,N,B_{\max})$ and schedule $s$ are specialized at compile time; $N$ denotes the local intermediate width after tensor-parallel sharding. Static dimensions allow the compiler to determine WGMMA loop bounds, TMA descriptors and shared-memory layouts precisely. These specializations are generated from a declarative shape table rather than handwritten: supporting a compatible model requires only registering its per-GPU shape and regenerating the kernel instances and dispatch entries, without changing the underlying MoE algorithm.

For each shape, MonoMoE precompiles schedules parameterized by $(\textsc{grid},G_{\mathrm{up}},K_{\mathrm{up}},G_{\mathrm{down}},K_{\mathrm{down}},D_{\mathrm{up}},D_{\mathrm{down}})$. Here, \textsc{grid} is the number of persistent CTAs; $G_{\mathrm{up}}$ and $G_{\mathrm{down}}$ are the numbers of CTAs that jointly process one expert in the two projections; $K_{\mathrm{up}}$ and $K_{\mathrm{down}}$ are the corresponding reduction-tile widths; and $D_{\mathrm{up}}$ and $D_{\mathrm{down}}$ are the weight-prefetch depths. The up and down group sizes and prefetch depths are selected independently, while readiness flags connect the two projection schedules. A configuration enumerator retains only candidates satisfying tensor-core divisibility, CTA-group compatibility, grid-residency, and shared-memory-capacity constraints. The candidates are then tuned offline: configurations that fail numerical validation are rejected, and the remainder are ranked by CUDA-Graph latency over the target token-count range and representative routing distributions. Serving selects the fixed configuration associated with the model shape and GPU through a generated dispatch table.

\section{Evaluation}
\subsection{Experimental Setup}
\label{eval:setup}
\paragraph{Hardware and software.}
We implement MonoMoE in CUDA and integrate it into vLLM~\cite{kwon2023vllm}. We evaluate on NVIDIA H200 GPUs~\cite{nvidia2024h200}, each with 132 SMs, 141\,GB of HBM3e, and 4.8\,TB/s peak memory bandwidth. All MoE weights and activations use E4M3 FP8. Weights use $128\times128$ block scales, while input activations are quantized inside MonoMoE at 128-element granularity. Both MonoMoE and the baseline execute under vLLM's CUDA-Graph decode path.

\paragraph{Models and workload.}
Table~\ref{tab:eval-config} covers four model configurations and substantially different per-GPU projection shapes. Here, $H$ is the hidden dimension and $N$ is the local expert intermediate width after TP sharding. All models have 256 routed experts and select eight experts per token. Qwen uses softmax routing; DeepSeek-V3.1~\cite{deepseekai2024v3} and GLM-5.2 use sigmoid routing with expert-selection bias and a routed scaling factor of 2.5. The Qwen experiments use one GPU, while DeepSeek-V3.1 and GLM-5.2 use eight-way tensor parallelism (TP).

\paragraph{Baselines.}
Our first baseline is vLLM's production Triton grouped-GEMM implementation~\cite{kwon2023vllm}, \texttt{fused\_moe}, which we refer to as \emph{vLLM Grouped GEMM}. It includes top-$k$ routing, token alignment, activation quantization, the two grouped GEMMs, SwiGLU, requantization, and final reduction. Thus, the comparison covers the same operator boundary as MonoMoE rather than only the two expert projections. For a fair comparison, we use vLLM's shape-specific configurations tuned for optimal \texttt{fused\_moe} performance. Batched 3D GEMM is primarily optimized for long-sequence prefill, where many tokens are assigned to each expert and the cost of constructing a padded expert-major tensor can be amortized over large GEMMs. Therefore, we use grouped GEMM as the baseline for our decode-focused evaluation. This choice also reflects current production practice: serving systems such as vLLM use the grouped-GEMM path for low-token MoE decode.

We also evaluate \emph{FlashMoE-FP8}, an FP8 adaptation of FlashMoE's persistent, token-major MoE kernel~\cite{aimuyo2025flashmoe}, using the same per-GPU model shapes and routing policies. To isolate its kernel runtime, we disable FlashMoE's token-dispatch and combine components, add FP8 quantization, and retain its routing and expert-computation stages. The reported FlashMoE-FP8 numbers therefore characterize this retained kernel path rather than the complete distributed dispatch-to-combine pipeline.

\paragraph{MonoMoE configurations.}
For each model shape, Table~\ref{tab:eval-config} reports the fixed configuration selected by offline tuning over the seven schedule parameters defined in Section~\ref{design:tuning}. Every CTA has 384 threads, with eight calculation warps and four prefetch/epilogue warps.

\begin{table*}[t]
\centering
\caption{Evaluated block-wise FP8 model configurations and default MonoMoE schedules. Dimensions are per GPU; $N$ is the local intermediate width after TP sharding. $G_{\mathrm{up}}$ and $G_{\mathrm{down}}$ are CTAs per expert, while $D_{\mathrm{up}}$ and $D_{\mathrm{down}}$ are weight-prefetch depths. All models use $E{=}256$ experts and top-$k{=}8$, and all CTAs use 384 threads.}
\label{tab:eval-config}
\small
\setlength{\tabcolsep}{3pt}
\begin{tabular}{lrrrlrrrrrrr}
\toprule
& \multicolumn{4}{c}{Model configuration}
& \multicolumn{7}{c}{MonoMoE schedule} \\
\cmidrule(lr){2-5}\cmidrule(lr){6-12}
Model & TP & $H$ & $N$ & Routing
& \textsc{grid} & $G_{\mathrm{up}}$ & $K_{\mathrm{up}}$
& $G_{\mathrm{down}}$ & $K_{\mathrm{down}}$ & $D_{\mathrm{up}}$
& $D_{\mathrm{down}}$ \\
\midrule
Qwen3.5-35B-A3B
& 1 & 2048 & 512 & softmax
& 128 & 8 & 256 & 8 & 256 & 4 & 2 \\
Qwen3.5-122B-A10B
& 1 & 3072 & 1024 & softmax
& 128 & 8 & 256 & 8 & 128 & 2 & 2 \\
GLM-5.2
& 8 & 6144 & 256 & sigmoid + bias
& 128 & 2 & 256 & 16 & 128 & 2 & 2 \\
DeepSeek-V3.1
& 8 & 7168 & 256 & sigmoid + bias
& 128 & 2 & 256 & 28 & 128 & 2 & 2 \\
\bottomrule
\end{tabular}
\end{table*}

\paragraph{Metrics and methodology.}
For isolated-kernel experiments, we generate FP8 expert weights, BF16 inputs, and router logits for $B\in\{1,2,4,8\}$. Latency is averaged over 200 CUDA-Graph replays after 20 warmup iterations. For end-to-end performance evaluation, we use vLLM's built-in fixed-shape Sonnet text workload~\cite{kwon2023vllm}: 50 prompts with 1,600 input tokens, 200 generated tokens, and a 200-token shared prefix. Prefix caching is disabled, chunked prefill is enabled, and request concurrency is swept over $\{1,2,4,8\}$ with a fixed seed. We report mean time per output token (TPOT), which averages post-first-token generation latency per request. For numerical validation, we compare the FP8 kernel output against an FP32 implementation of the same operator on the same randomly generated inputs. We also run end-to-end vLLM decoding on downstream tasks and compare task accuracy with the Triton baseline.

\subsection{Kernel-Level Performance}
\label{eval:kernel}

\begin{figure*}[t]
\centering
\includegraphics[width=\textwidth]{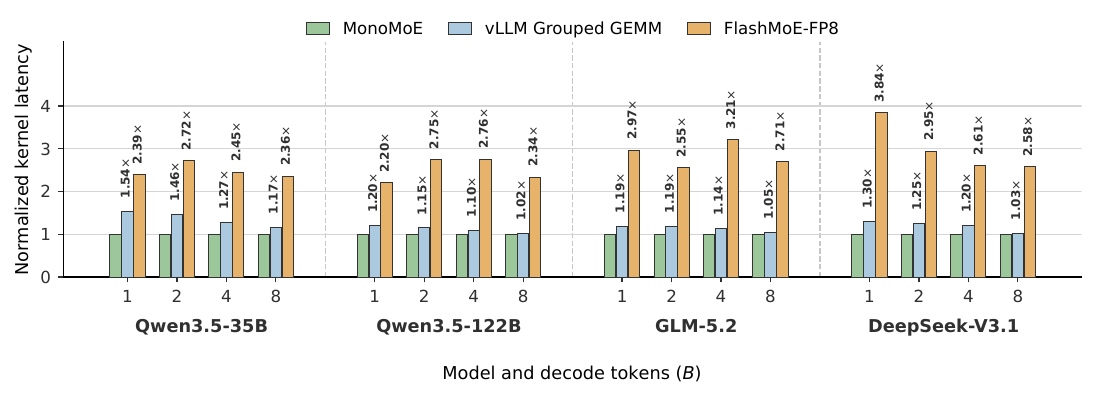}
\vspace{-2em}
\caption{Normalized isolated routed-MoE latency under synthetic routing. Each model and token-count group is normalized to MonoMoE at 1.0, with bars ordered as MonoMoE, vLLM Grouped GEMM, and FlashMoE-FP8. Labels report relative latency. Lower is better.}
\label{fig:kernel-latency}
\end{figure*}

Figure~\ref{fig:kernel-latency} summarizes the isolated-kernel measurements and normalizes each model and token-count group to MonoMoE's latency. MonoMoE, vLLM Grouped GEMM and FlashMoE-FP8 cover the same complete routed-MoE operator. Across $B\in\{1,2,4,8\}$, MonoMoE achieves speedups over vLLM Grouped GEMM of $1.17$--$1.54\times$ on Qwen3.5-35B, $1.02$--$1.20\times$ on Qwen3.5-122B, $1.05$--$1.19\times$ on GLM-5.2, and $1.03$--$1.30\times$ on DeepSeek-V3.1, using the tuned configurations reported in Table~\ref{tab:eval-config}. The advantage is largest at the most latency-sensitive token counts and narrows as additional tokens give the token-major baseline more GEMM tiles over which to amortize dispatch and pipeline transients. On the same normalized scale, FlashMoE-FP8 incurs $2.36$--$2.72\times$ MonoMoE's latency on Qwen3.5-35B, $2.20$--$2.76\times$ on Qwen3.5-122B, $2.55$--$3.22\times$ on GLM-5.2, and $2.58$--$3.84\times$ on DeepSeek-V3.1. FlashMoE-FP8 is therefore slower than MonoMoE at every evaluated point, by $2.20$--$3.84\times$ overall.

\paragraph{Matrix-level numerical accuracy.}
To evaluate matrix-level numerical accuracy, we draw FP32 activation and expert-weight matrices from both a standard Gaussian distribution and a heavy-tailed Student-$t$ distribution with three degrees of freedom. Each matrix is normalized to the same root-mean-square magnitude before applying the production BF16 and block-wise FP8 conversions, and routing assignments are held fixed across implementations. For each model shape and $B\in\{1,2,4,8\}$, we compare MonoMoE and vLLM Grouped GEMM against the same FP32 reference. Across all evaluated distributions, model shapes, and token counts, cosine similarity remains at least 0.998.

\subsection{Kernel Performance Analysis}
\label{eval:analysis}

The weight-major MonoMoE kernel directly addresses the three bottlenecks identified in Section~\ref{bottleneck}: token-major preprocessing, low effective memory throughput, and exposed auxiliary work. For Qwen3.5-35B-A3B-FP8 at $B\in\{1,2,4,8\}$, eliminating token-major padding and preprocessing removes 5.467, 5.335, 5.446, and 5.663\,$\mu$s, corresponding to 26.7\%, 24.0\%, 17.3\%, and 13.2\% of the respective kernel latencies of 20.44, 22.27, 31.53, and 43.01\,$\mu$s. We compute operator-level effective memory throughput for the Triton grouped-GEMM baseline as the execution-time-weighted mean of the per-kernel DRAM throughput. Across the same token counts, MonoMoE increases effective throughput from 10.62\%, 19.36\%, 30.78\%, and 45.07\% of peak to 21.20\%, 37.51\%, 49.06\%, and 57.99\%, corresponding to improvements of $2.00\times$, $1.94\times$, $1.59\times$, and $1.29\times$. MonoMoE further overlaps routing, quantization, activation, requantization, and partial accumulation with the expert-weight stream, reducing the auxiliary work exposed on the critical path. Collectively, these effects explain the kernel-level speedups in Figure~\ref{fig:kernel-latency} and map directly to the entitlement decomposition in Equation~\ref{eq:entitlement}: single-launch execution eliminates inter-kernel boundary gaps, while continuous weight streaming and auxiliary-work overlap recover the throughput component. The diminishing gains with increasing $B$ are consistent with Table~\ref{tab:entitlement}, which identifies the low-token regime as the biggest opportunity.

\subsection{End-to-End Serving}
\label{eval:e2e}

\begin{figure*}[t]
\centering
\includegraphics[width=\textwidth]{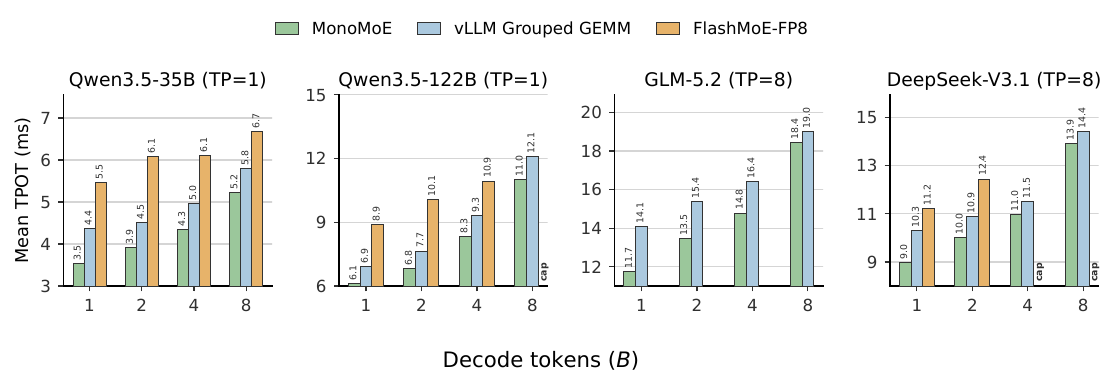}
\vspace{-2em}
\caption{Mean TPOT on the Sonnet workload across decode token counts. For these end-to-end results, FlashMoE-FP8 adds FP8 quantization to the original FlashMoE implementation while retaining dispatch and combine. Results are available for both Qwen models and the supported DeepSeek-V3.1 points. \emph{cap} marks configurations rejected at initialization because FlashMoE's estimated expert-plus-dispatch processor count exceeds the fixed 132-SM budget on H200. MonoMoE executes all shown shapes with tuned configurations. FlashMoE-FP8 cannot be integrated into vLLM's GLM-5.2 serving path, so the GLM-5.2 panel compares MonoMoE only with vLLM Grouped GEMM.}
\label{fig:e2e-tpot}
\end{figure*}

Figure~\ref{fig:e2e-tpot} reports paired runs from the same benchmark invocation. At $B\in\{1,2,4,8\}$, MonoMoE reduces TPOT relative to vLLM Grouped GEMM by $18.7\%$, $13.0\%$, $12.3\%$, and $9.9\%$ on Qwen3.5-35B. The corresponding reductions are $11.5\%$, $10.7\%$, $10.7\%$, and $9.1\%$ on Qwen3.5-122B, and $16.7\%$, $12.3\%$, $9.9\%$, and $2.9\%$ on GLM-5.2. The decline is most pronounced for GLM-5.2 because its TP=8 serving path includes communication and non-MoE work that the local kernel does not reduce. On DeepSeek-V3.1, MonoMoE records TPOTs of 8.98, 10.00, 10.96, and 13.90\,ms, compared with 10.3, 10.9, 11.5, and 14.4\,ms for Grouped GEMM. These correspond to TPOT reductions of $12.8\%$, $8.3\%$, $4.7\%$, and $3.5\%$. Thus, communication and non-MoE work in the eight-way TP serving path increasingly dilute the local-kernel gain as concurrency grows.

MonoMoE outperforms FlashMoE-FP8 at every supported end-to-end point. For this comparison, FlashMoE-FP8 adds FP8 quantization to the original FlashMoE implementation while retaining its complete dispatch, expert-computation, and combine path. MonoMoE reduces TPOT by $35.1\%$, $35.5\%$, $28.9\%$, and $21.8\%$ on Qwen3.5-35B; by $31.4\%$, $32.2\%$, and $24.0\%$ on Qwen3.5-122B at $B\in\{1,2,4\}$; and by $20.0\%$ and $19.6\%$ on DeepSeek-V3.1 at $B\in\{1,2\}$. FlashMoE-FP8 reports \emph{cap} for Qwen3.5-122B at $B=8$ and DeepSeek-V3.1 at $B\in\{4,8\}$. FlashMoE-FP8 cannot be integrated into vLLM's GLM-5.2 serving path, so no end-to-end result is available for that model.

The \emph{cap} labels reflect FlashMoE's scheduling capacity rather than an out-of-memory condition or a general restriction on model size. FlashMoE reserves one CTA for its on-device scheduler and estimates the processor CTAs required for expert computation and token dispatch from the per-rank token count $B$ and the model and routing parameters $(H,N,E,k)$. Initialization fails when the scheduler and processor CTAs together exceed H200's 132-SM budget; increasing the number of local experts, tokens per rank, or hidden width can therefore trigger the limit. MonoMoE instead repartitions a fixed persistent grid over expert-weight tiles through generated shape-specific schedules, allowing it to execute every evaluated model shape and token count.


\subsection{End-to-End Accuracy}
\label{eval:accuracy}

We evaluate deterministic end-to-end generation on GSM8K~\cite{cobbe2021gsm8k}, a grade-school mathematical-reasoning benchmark (5-shot, flexible-extract exact match), and HumanEval~\cite{chen2021humaneval}, a Python code-generation benchmark (0-shot, pass@1). Table~\ref{tab:e2e-accuracy} reports the mean score and binomial standard error. We define accuracy loss as the Grouped GEMM mean minus the MonoMoE mean. A value of zero means that the two backends achieve the same score, while a negative value means that MonoMoE scores higher. Across all models and both tasks, MonoMoE preserves the serving accuracy of Grouped GEMM: whenever MonoMoE's mean score is lower, the difference is smaller than the combined standard error, while all remaining results indicate parity or favor MonoMoE. Thus, replacing the baseline MoE path with MonoMoE introduces no measurable accuracy degradation through the complete serving stack.

\begin{table}[t]
\centering
\caption{End-to-end downstream-task accuracy at batch size one, reported as
mean $\pm$ binomial standard error. Accuracy loss is the difference in means,
Grouped GEMM minus MonoMoE, so lower is better.}
\label{tab:e2e-accuracy}
\setlength{\tabcolsep}{3pt}
\small
\begin{tabular}{@{}llccc@{}}
\toprule
Model & Task & Grouped G. & MonoMoE & Loss \\
\midrule
Qwen3.5-35B   & GSM8K     & $.750{\pm}.012$ & $.790{\pm}.011$ & $-.040$ \\
              & HumanEval & $.890{\pm}.024$ & $.870{\pm}.026$ & $+.020$ \\
\midrule
Qwen3.5-122B  & GSM8K     & $.868{\pm}.009$ & $.866{\pm}.009$ & $+.002$ \\
              & HumanEval & $.854{\pm}.028$ & $.854{\pm}.028$ & $.000$ \\
\midrule
GLM-5.2       & GSM8K     & $.952{\pm}.006$ & $.948{\pm}.006$ & $+.004$ \\
              & HumanEval & $.677{\pm}.037$ & $.720{\pm}.035$ & $-.043$ \\
\midrule
DeepSeek-V3.1 & GSM8K     & $.953{\pm}.006$ & $.959{\pm}.005$ & $-.006$ \\
              & HumanEval & $.732{\pm}.035$ & $.768{\pm}.033$ & $-.037$ \\
\bottomrule
\end{tabular}
\end{table}

\section{Discussion} 

\subsection{Target Operating Regime}
\label{discussion:regime}

MonoMoE targets memory-bound autoregressive decode with a small number of concurrently processed tokens per GPU, denoted by $B$; this token count is distinct from the accumulated context length of a request. Its strongest regime is $B\leq8$, where the complete token tile fits the smallest fine-grained tensor-core $N$ shape, while token-major grouped GEMM commonly allocates an underfilled $M$ tile for each active expert. The baseline consequently exposes little token-axis parallelism, short-lived grids, and poorly amortized preprocessing. MonoMoE instead derives CTA-level parallelism from expert-weight tiles and keeps those CTAs resident across experts and projections.

As $B$ grows, additional routed tokens fill token-major $M$ tiles and amortize token sorting, launch boundaries, and pipeline fill/drain costs. Accordingly, the isolated-kernel speedup across the four models decreases from $1.19$--$1.54\times$ at $B=1$ to $1.02$--$1.17\times$ at $B=8$ (Figure~\ref{fig:kernel-latency}), while the end-to-end TPOT reduction decreases from $11.5$--$18.7\%$ to $2.9$--$9.9\%$ (Figure~\ref{fig:e2e-tpot}); communication and non-MoE work further dilute the TP=8 results. FlashMoE-FP8 remains slower at every isolated-kernel point. A longer decode context alone does not increase $B$, whereas prefill and chunked prefill process many sequence positions together and provide enough token-axis parallelism to amortize dispatch. MonoMoE therefore targets decode steps with $B\leq8$; for larger token sets, grouped or batched GEMM becomes increasingly appropriate.

\subsection{Communication in the MonoMoE}

MonoMoE fuses expert computation into one persistent launch, but a partitioned MoE layer still requires distributed communication: TP all-reduces the layer output, while EP dispatches tokens to expert owners and returns their results~\cite{lepikhin2021gshard,rajbhandari2022deepspeedmoe,aimuyo2025flashmoe}. Folding these collectives into the kernel does not improve single-node, small-batch decode because NVLink transfers the small messages in only a few microseconds, leaving cross-rank synchronization as the dominant exposed cost. A waiting persistent kernel also keeps one CTA resident on every SM, causing the grid to spin while communication traffic competes with the bandwidth-bound expert-weight stream. TP offers no useful overlap window because its all-reduce is terminal, and MonoMoE's \texttt{atomicAdd}-based accumulation does not produce an output element until all contributing expert groups finish. Exposing partial results earlier would require per-group buffers, roughly doubling output traffic and adding a reduction pass, while the receive would still wait for the slowest rank. Separate TP collectives are therefore preferable in MonoMoE's target regime; fusion becomes more attractive at higher TP degrees or during prefill, where larger transfers provide meaningful overlap opportunities.

For EP, we implemented fused dispatch and combine using symmetric peer memory: ranks stage tokens and partial outputs in peer-mapped buffers, and the kernel peer-reads remote activations during routing and sums the rows owned by each rank, replacing the corresponding all-gather and reduce-scatter operations. This implementation is correct but performs at parity because the short NVLink transfer leaves the intrinsic rank-to-rank wait unchanged, while the shared expert---the only independent computation---is already overlapped by the runtime. The tradeoff changes at larger batch sizes, higher EP degrees, or across scale-out interconnects, where larger payloads and longer expert GEMMs create a useful overlap window. In those regimes, a communication-overlapping megakernel could process early-arriving experts while later tokens remain in flight; for single-node, small-batch decode, separate communication and fused expert computation remain the better organization.


\bibliography{example_paper}
\bibliographystyle{mlsys2025}



\end{document}